\documentclass[a4paper,fleqn,usenatbib]{mnras}

\usepackage{newtxtext,newtxmath}
\usepackage[T1]{fontenc}
\usepackage{ae,aecompl}

\usepackage{graphicx}	% Including figure files
\usepackage{amsmath}	% Advanced maths commands
\usepackage{amssymb}	% Extra maths symbols
\newcommand{\bjdtdb}{\ensuremath{\rm {BJD}}}
\newcommand{\feh}{\ensuremath{\left[{\rm Fe}/{\rm H}\right]}}
\newcommand{\teff}{\ensuremath{T_{\rm eff}}}
\newcommand{\logg}{\ensuremath{\log \rm g}}

\newcommand{\msun}{\ensuremath{\,M_\odot}}
\newcommand{\rsun}{\ensuremath{\,R_\odot}}
\newcommand{\lsun}{\ensuremath{\,L_\odot}}
\newcommand{\mj}{\ensuremath{\,M_{\rm J}}}
\newcommand{\rj}{\ensuremath{\,R_{\rm J}}}
\newcommand{\fave}{\langle F \rangle}
\newcommand{\fluxcgs}{10$^9$ erg s$^{-1}$ cm$^{-2}$}

\newcommand{\revision}{\textcolor{black}}

\title[A modified CoRoT detrend algorithm]{A modified CoRoT detrend algorithm and the discovery of a new planetary companion}

\author[R.~C.~Boufleur et al.]{
Rodrigo C. Boufleur,$^{1}$\thanks{E-mail: rcboufleur@on.br}
Marcelo Emilio,$^{2,3}$
Eduardo Janot-Pacheco,$^{4}$\newauthor
~Laerte Andrade,$^{2}$
Sylvio Ferraz-Mello,$^{4}$
Jos\'e-Dias do Nascimento J\'unior,$^{5,6}$\newauthor
~J. Ramiro de La Reza$^{1}$
\\
$^{1}$Observat\'{o}rio Nacional - MCTIC, Rio de Janeiro, RJ, Brazil\\
$^{2}$Universidade Estadual de Ponta Grossa, Ponta Grossa, PR, Brazil\\
$^{3}$Institute for Astronomy, University of Hawaii, USA\\
$^{4}$Instituto de Astronomia, Geof\'{i}sica e Ci\^{e}ncias Atmosf\'{e}ricas - USP, S\~{a}o Paulo, SP, Brazil\\
$^{5}$Univ. Federal do Rio Grande do Norte, UFRN, Dep. de F\'{i}sica, CP 1641, 59072-970, Natal, RN, Brazil\\
$^{6}$Harvard-Smithsonian Center for Astrophysics, 60 Garden St. Cambridge, MA 02138, USA
}

\date{Accepted XXX. Received YYY; in original form ZZZ}

\pubyear{2017}

\begin{document}
\label{firstpage}
\pagerange{\pageref{firstpage}--\pageref{lastpage}}
\maketitle

\begin{abstract}
%This is a simple template for authors to write new MNRAS papers.
%The abstract should briefly describe the aims, methods, and main results of the paper.
%It should be a single paragraph not more than 250 words (200 words for Letters).
%No references should appear in the abstract.
We present MCDA, a modification of the CoRoT detrend algorithm (CDA) suitable to detrend chromatic light curves. By means of robust statistics and better handling of short term variability, the implementation decreases the systematic light curve variations and improves the detection of exoplanets when compared with the original algorithm. All CoRoT chromatic light curves (a total of 65,655) were analysed with our algorithm. Dozens of new transit candidates and all previously known CoRoT exoplanets were rediscovered in those light curves using a box-fitting algorithm. For three of the new cases spectroscopic measurements of the candidates' host stars were retrieved from the ESO Science Archive Facility and used to calculate stellar parameters and, in the best cases, radial velocities. In addition to our improved detrend technique we announce the discovery of a planet that orbits a $0.79_{-0.09}^{+0.08}$\rsun~star with a period of $6.71837\pm0.00001$~days and has $0.57_{-0.05}^{+0.06}$\rj~and $0.15\pm0.10$\mj. We also present the analysis of two cases in which parameters found suggest the existence of possible planetary companions.

\end{abstract}


%%%%%%%%%%%%%%%%%%%%%%%%%%%%%%%%%%%%%%%%%%%%%%%%%%

%%%%%%%%%%%%%%%%% BODY OF PAPER %%%%%%%%%%%%%%%%%%

\section{Introduction}

Stellar light curves may carry intrinsically long term trends which are associated to the variability in the stellar flux itself. Other sources of trends appear from space-based photometry measurements as systematic variations requiring special analysis to optimize, for instance, planetary transit detections. Earth-like planets demand high quality measurements since the transit generates variations in the flux measurements of, at most, a few percent. Moreover, in order to perform the best assessment of the data we must be able to deal in the best possible way with the presence of intrinsic stellar variability as well as systematic trends caused by the influence of external error sources. Usually detrend techniques are performed combining mathematical and statistical operations, and their design is closely related to the chosen detection method as discussed by \cite{moutou2005}, \cite{aigrain2004} and \cite{mislis2010}. For a better understanding of light curves detrending problems see \cite{mazeh2007}, \cite{ofir2010}, \cite{kovacs2008} and \cite{kim2009}.

CoRoT\footnote{The CoRoT space mission, launched on 2006 December 27, was developed and is operated by the CNES, with participation of the Science Programs of ESA, ESA's RSSD, Austria, Belgium, Brazil, Germany and Spain.} (COnvection ROtation and planetary Transits) was a space-based mission focused on precise photometric measurements, allowing both stellar seismology analysis and the detection of new exoplanets by means of the transit method \citep{auvergne2009}. It consisted of a $27$ cm telescope equipped with four CCDs cameras in two different fields, one for each main scientific goal, seismology channel (bright stars) and exoplanet channel (faint stars). The CoRoT public data set is composed of tens of thousands of light curves obtained in the chromatic or monochromatic mode and was assembled from January, 2007 to November, 2012 by means of short (SR) and long runs (LR). The pointing targets in this mission were chosen within two main regions in the sky, one towards the galactic center (LRc or SRc, $ \alpha = 18^{\text{h}}50^{\text{m}}$) and the other towards the galactic anti-center (LRa or SRa, $ \alpha = 06^{\text{h}}50^{\text{m}} $). Most stars monitored in the faint stars field were primarily sampled at a 512~s rate. Up to 2,000 stars could also be monitored at a 32~s rate when triggered the so called ``alarm mode'' \citep{Surace2008,Bonomo2012}.  The original field coverage comprised in the exoplanet channel during each pointing was about $1.4 \times 2.8$~square degrees, allowing CoRoT to observe up to 12,000 stars per run.

In this work we present a modification to the CoRoT detrend algorithm  (CDA) \cite{mislis2010}. Our detrend filtering technique focuses on the treatment of sudden statistical fluctuations, a feature that strongly affects folding phase transit search methods. \citet{mislis2010} found for the chromatic (CHR) light curves in CoRoT Initial Run IRa01 that less than 1\% of this type of event appear in the three colour channels simultaneously. So, this technique uses a comparison obtained in each colour channel individually to attenuate the presence of discontinuities.

All CHR CoRoT light curves --- 65,655 in total --- were analysed in this work with our algorithm. All CoRoT previously known exoplanets in those light curves were found and new candidates appeared. In section \ref{sec:Algorithm}  we describe our algorithm, its performance and validation. Section \ref{sec:photometry_analysis} describes the photometric data, the transit detections and modeling, and section  \ref{sec:spectra} the spectroscopy analysis for the found new candidates. The results are presented in section \ref{sec:results} and our conclusions in section \ref{sec:conclusions}.

\section{Algorithm}
\label{sec:Algorithm}

    \subsection{Description}
    \label{sec:alg_description} % used for referring to this section from elsewhere

\citet{mislis2010} developed a technique to detrend CoRoT light curves from sudden intensity changes. Such changes can be of stellar activity origin, although the vast majority are clearly instrumental in nature \citep{pinheiro2008} and strongly affect folding phase transit search methods. Although space based telescopes can provide high precision stellar measurements due to the lack of atmosphere, there are other sources of error to deal with. The measurements provided by the CoRoT satellite detectors frequently suffer with the presence of spurious charge added to the integrated stellar flux as estimated by \cite{mislis2010}. It is primarily the result of the impact of highly energized particles that overcame the protective shielding onto the detectors. \cite{pinheiro2008} showed that in the CoRoT case this phenomenon is closely related to the satellite's low Earth orbit since these particles are originated from the interactions of cosmic rays in the Earth's upper atmosphere. Indeed, the passage over the South Atlantic Anomaly produces the highest rate of radiation. These sudden variations in the data, commonly referred to as ``jumps'' in the literature, can be a transitory damage,  suffering decay after a while, or they can generate a permanent bright pixel when the impact alters fundamental properties of the sensor.

Our main goal is to improve the algorithm sensibility in the presence of discontinuities. Therefore, our main modifications of the CDA algorithm are:
\\
1. We replaced the originally employed third order polynomial fit with a resistant moving average.
 So, the mean value of the light curve section is taken with respect to its central value with an equal number of measurements on either sides,
        
        \begin{equation}
            \label{eq:ma}
            \overline{MA} = \frac{x_m + x_{m-1} + \cdots + x_{m-(n-1)}}{n}\text{,}
        \end{equation}
    where $n$ is an arbitrary value with respect to the sampling. At this point care is important to avoid overlapping the boxcar size with the typical transit durations as well as to minimize attenuation. In the case of the CoRoT data one expects transits up to around 5 hours. Additionally, high frequency signals are not filtered out. 
\\
2. In order to deal with the most significant \emph{jumps} and to minimize the local statistical error propagation (that are method dependent) in the final results, we implemented a robust statistics rather than the one adopted by \cite{mislis2010}. At first, we separated the measurements of the three channels into three independent light curves. This colour decomposition, here informally designated as red, green and blue channels (R,G,B), was obtained reading out the split regions within each stellar mask on the CCD, according to the spread of the white light. Each set was then divided into subsets with one day length. A normalization using the resistant mean for each subset $i$ was performed
        
        \begin{equation}
            [R',G',B']_{i} = \frac{[R,G,B]_i}{\overline{x}_{res_{[R,G,B]}}}.
        \end{equation}
        
        For each colour set we randomly selected five subsets and computed the mean resistant standard deviation
        
        \begin{equation}
            \overline{\sigma}_{res_{[R',G',B']}} = \frac{1}{5}\sum_{i=1}^{5}\sqrt{\frac{1}{l-1}\sum_{j=k_i}^{k_i+l}{\left([R',G',B']_{j} - \overline{x}_{res_{[R,G,B]_{i}}}\right)^2}},
        \end{equation}
        
    where  $i$ is regarded to each subset, $j$ to each element within the subsets and $l$ to the number of elements in the subset.
    The normalised subsets allows to do direct comparisons regarding statistical inference among the three channels. Also, we need to assess the significance of each $\overline{\sigma}_{res}$ relative to the light curve as a whole, i.e., estimate the presence of considerable statistical variation on each colour. To do so we first compute the resistant standard deviation of each colour set
        
        \begin{equation}
            \sigma_{res_{[R',G',B']}} = \sqrt{ \frac{1}{n-1} \sum_{i=1}^{n}{\left( [R',G',B']_i - \overline{x}_{res_{[R,G,B]}} \right)^2} },
        \end{equation}
        
        where $n$ is the set of elements that lie within the sigma cut established in our robust analysis. We compute then the relative deviations of each colour set to determine in which set the highest deviation is located,
        
        \begin{equation}
            \sigma_{rel_{[R',G',B']}} = \frac{\sigma_{res_{[R',G',B']}}}{\overline{\sigma}_{res_{[R',G',B']}}}.
        \end{equation}
        
        Finally, the relative deviations in each colour are ranked and the channel with the highest deviation is replaced by the average value of the channels with the highest and lowest deviations as follows,
        
        \begin{equation}
            C_{corr} = \frac{ C_{high_{\sigma_{rel}}} + C_{low_{\sigma_{rel}}} }{2}.
        \end{equation}
        
        This procedure is repeated in a loop while the relative deviations decrease relative to the former ones. After that, no significant statistical information can be used to reduce the presence of jumps. 
        
        %light curve with jumps plot
        \begin{figure*}
        	\includegraphics[width=\textwidth]{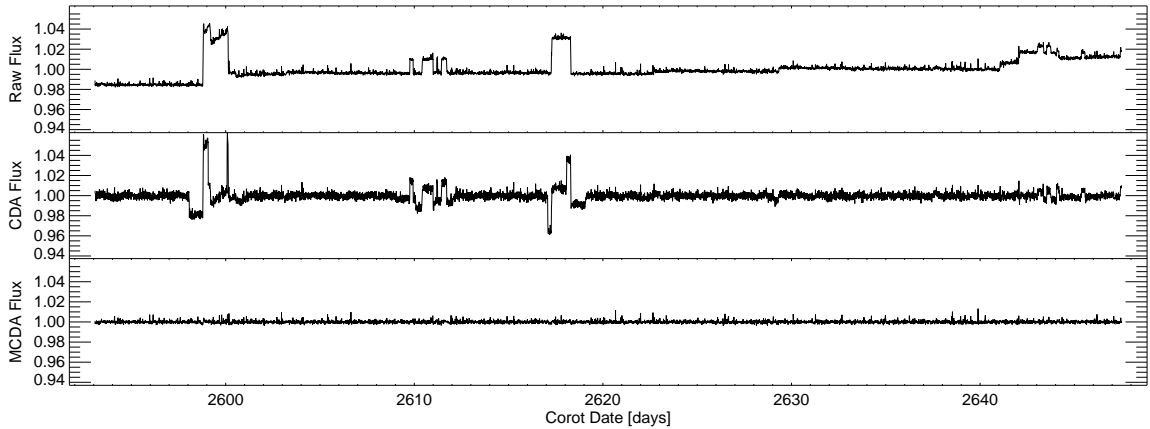}
            \caption{Comparison between the original flux observed with the CoRoT satellite and the results of the applied detrends methods discussed in section \ref{sec:Algorithm}. At the top a CoRoT light curve shows the presence of discontinuities in the original normalised flux. The middle shows the light curve detrended using the CDA algorithm. The bottom shows the correction using the MCDA method described in this work.}
            \label{fig:cdacdam}
        \end{figure*}

    \subsection{Performance and validation}
    \label{sec:alg_performance}

        \subsubsection{Application to real data}
    
            We show now the performance of our technique using the data provided by the CoRoT mission. 
            The light curves were obtained from the \href{idoc-corot.ias.u-psud.fr}{IAS CoRoT Public Archive}. The data we used from the initial runs were processed with pipeline version 1.x and more recent ones were already available with pipeline version 2.x.
            
            For the sake of simplicity we call our algorithm MCDA (Modified CoRoT Detrend Algorithm). Our analysis is restricted to the CHR light curves, that were obtained using a dispersive device in front of the exoplanet channel detectors, which allowed to discriminate transits events (quasi-achromatic) from stellar activity (strongly chromatic), among other characteristics \citep{Auvergne2003}.
            
            To do a confident assessment on how much we were able to reduce the dispersion on the data due to the presence of discontinuities we performed the median of absolute deviations from the median (MAD) \citep{Hoaglin2000}, which is a robust estimator of statistical dispersion. Using these estimates we are able to assess qualitatively and quantitatively how well our detrending method performs when our results are compared to the raw measurements as well as to the detrending performed by the CDA. 
            
            In figure \ref{fig:cdacdam} we show an arbitrary light curve from the CoRoT data set with the presence of \emph{jumps}. The original flux suffers as much from sudden displacements as from permanent deviations in the measurements. CDA turns out to be very effective to deal with long trend deviations, but less so with short term variations. It is probably the effect of a poor performance of the polynomial fit at the sharp changes in the vicinity of the \emph{jumps}. On the other hand, MCDA was able to remove all the short and long term discontinuities as it can be seen in the bottom plot of the figure. The effectiveness of the technique, however, depends heavily on former assumptions as, for instance, the parameters adopted to perform the smoothing of the light curve (Eq. \ref{eq:ma}).
            
\begin{figure}
        	    \includegraphics[angle=270,width=\columnwidth]{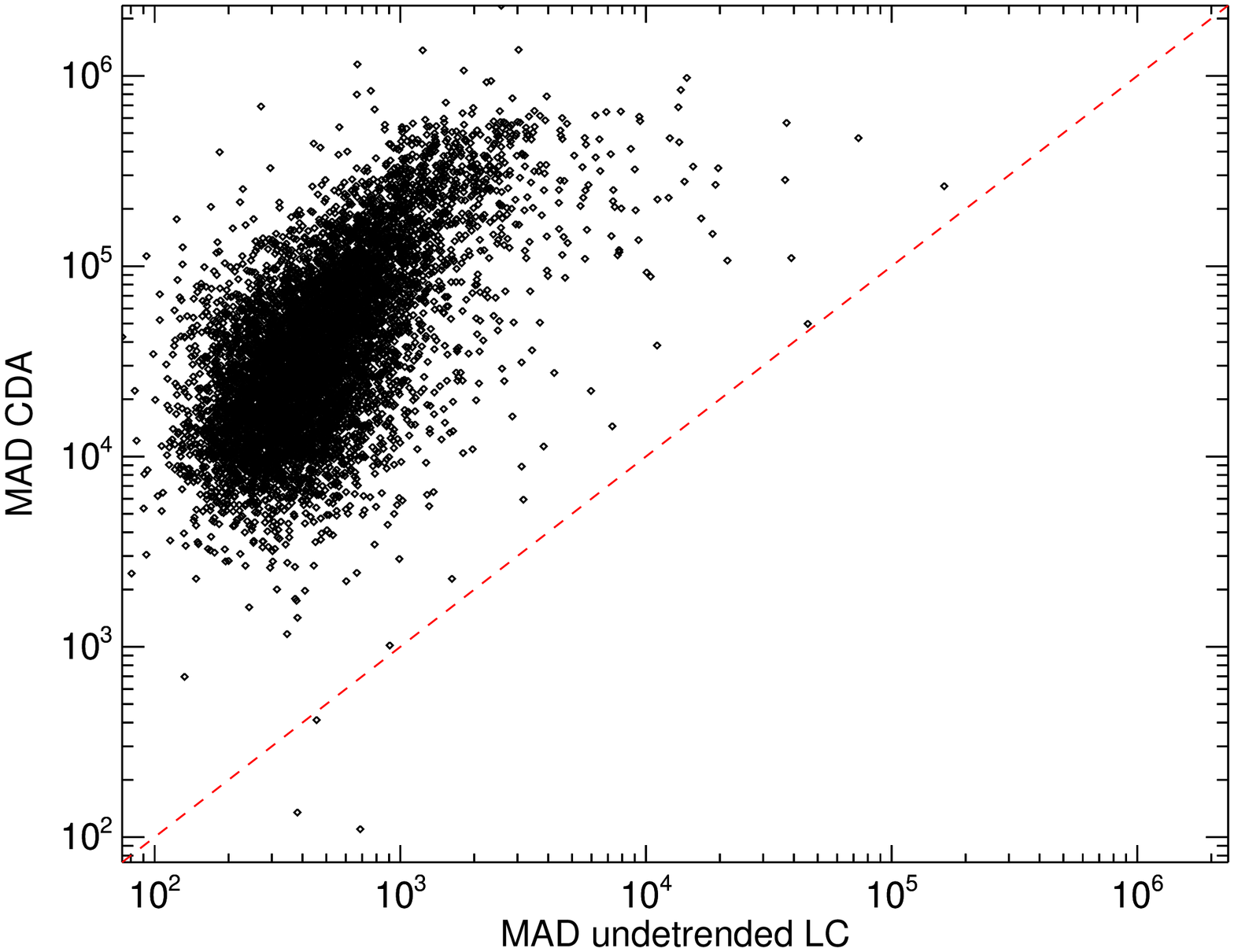}
        	    \includegraphics[angle=270,width=\columnwidth]{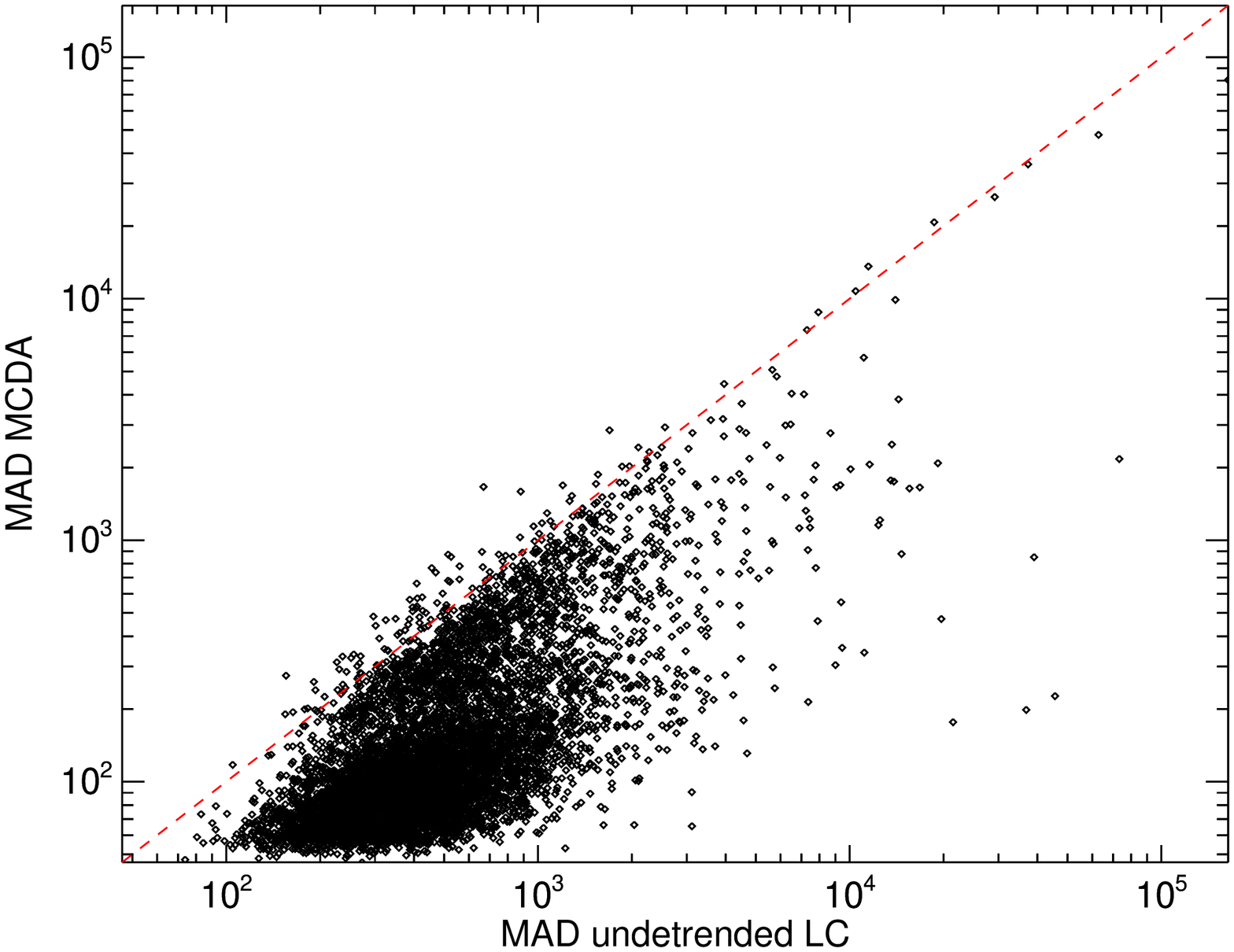}
                \caption{In the top figure we show the median of absolute deviations from the median (MAD) of each light curve for the LRa01 run after detrending the data using the CDA described in \citet{mislis2010} against the MAD of the raw data set. The bottom figure exhibits the MAD results using our method. It is noticeable that the MCDA provides a higher reduction in the statistical dispersion originated from jumps and stellar variability. }
            \label{fig:plotmad}
            \end{figure}

            Of all runs available, LRa01 is especially interesting for its number of targets and also for had been widely studied to determine the noise properties of the mission's data \citep{Aigrain2009}. So, to compare the methods and test their effectiveness, we show in Fig.~\ref{fig:plotmad} the MAD of the light curves after applying CDA versus the MAD of the raw data, and the same for MCDA.             
            It is easy to see that MCDA was able to lower the MAD to almost 100\% of the 7,470 CHR targets. In a few cases it seems to have added a little noise to the light curves.

            The use of a robust approach to assess the deviation's statistics allows MCDA to handle much better the global dispersion of the data, assuring to a higher degree that the chosen sets to be corrected indeed suffer from \emph{jumps}. On the other hand, the lower performance of CDA comes from the way trends are filtered out, such as stellar variability, in combination with the fact that discontinuities in the data deviates the probability distribution from normality, making standard statistical estimators less sensitive.  

            All the 24 runs studied showed a high improvement on the reduction of the dispersion similar to the run LRa01 in comparison to the original light curves. This shows that the technique is indeed contributing to enhance the detections in the light curves. These transits would not be prominent before due to the dominant signal of the jumps.
            
            Since the method's procedure for correction is based on averaging the most affected colour with the least affected channel, a little increase on the occurrence of false positives could be expected, like signals that first appeared only in a single colour being imprinted now in other channels. This is however perfectly manageable since modeling of the candidates is preferably done with the raw light curves, where such false positives can be ruled out. On the other hand, the technique helps to lower the occurrence of false negatives, especially the cases where the transits are shallow and would be missed if the discontinuities were not properly handled in advance.

            %lra01 mad cda cdam plot

\section{Photometric Analysis}
\label{sec:photometry_analysis}
    \subsection{Transit detections}
    \label{sec:transits}
        Transits are caused by the passage of an opaque body in front of a star in the line of sight of the observer. Planetary transits themselves have usually the shape of a box with soft edges caused by the limb darkening observed on the stellar disk. \cite{rauer2007} and \cite{winn2010} explain in details the phenomenon of transits and the major misinterpretations that can originate from geometry or shape as well as from mimic situations such as background diluted binaries. Although this is a periodic event and there are many approaches to be employed in the frequency domain to detect these signals, box shaped algorithms such as Box-fitting Least Squares (BLS) \citep{kovacs2002} are more reliable to deal with low signal-to-noise ratio and also more efficient to perform transit searches as shown by \citet{aigrain2004}.

        CoRoT Exo field has in total 176,492 light curves. From those, 65,655 are chromatic and 110,837 are monochromatic. We performed the detrending in the entire CHR data set. After applying MCDA we did the search for periodic transits using the BLS algorithm. The output provided by the BLS algorithm is a periodogram. As defined by \cite{kovacs2002}, each spike in the periodogram has a signal detection efficiency (SDE), which is dependent on the number of frequencies used to test for the presence of box-shaped signals. We established a SDE cut and ranked our first generation candidates. 

        Many false positives can be ruled out at this stage. We tested for each candidate the difference in the depths of odd and even transits in the three channels independently. A visual inspection was carried out on each remaining pre-candidate checking for the occurrence of the transits in the three channels using the raw data. We also checked for the presence of shallow occultations and strong deep V-shaped transits, that are usually featured in binaries, due to the low radius ratio between the companions. To estimate the transit parameters in an optimal way we analysed the singular spectrum of the variability present in the measurements as explained next.
        
        Besides the already published CoRoT planets, we detected 45 still unpublished candidates within the depth limit of $2$\%. Radial velocity measurements were publicly available in the ESO Science Archive Facility for 11 of these cases, from which we chose three promising cases to present our findings.
    
    \subsection{Variability Modeling}
    \label{sec:ssa}
        It is well known and expected that detrend techniques introduce distortions in the transit shapes compromising in some level the information on depth and duration of the event, and consequently the determination of transit parameters. To make the best use of the information we performed a custom detrend in each candidate light curve with the aim of disentangling the transit signal from the variability present in the measurements. 
    
        The Singular Spectrum Analysis (SSA) is a technique based on the division of a time series into its constituent parts of trends, periodicity and residual structures. It does not assume any parametric model and has been well employed in studies of non trivial periodicities in Climate and Astronomy time series. To this end it makes use of the decomposition of the original time series into sub data sets, building a multidimensional series from which the singular spectrum is produced -- the spectral decomposition of the multidimensional series into its eigenvalues set \citep{elsner1996,ghil2002}.
        
        %first app of transit signal    
        To isolate the variability present in the light curve we modeled the undetrended light curve without the transits. To remove properly the transits signals we did the following. First, corrections were performed when heavy discontinuities understood as systematics were found. The data was then re-sampled to a 512~s cadence and long trends were treated using a Savitzky-Golay filter \citep{savitzky1964}. A trapezoidal function was fitted over the time series folded on the candidate's period to derive the planetary parameters using the equations derived by \cite{seager2003} and fitted with a Levenberg-Marquardt algorithm \citep{markwardt2009}. Finally, using the information on duration and epoch for each transit, they were located and removed from the corrected raw light curve and the gaps were interpolated using a polynomial function.
        
\begin{figure}
        	\includegraphics[angle=270,width=\columnwidth]{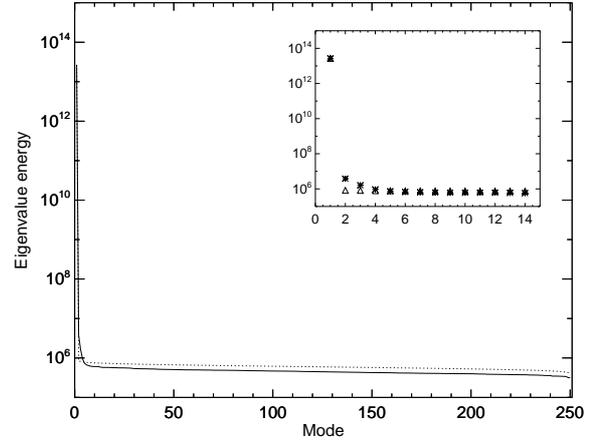}
        	\caption{Singular spectrum analysis of CoRoT ID 652345526. The filled line shows the eigenvalues of the 250 modes calculated using the corrected light curve. The dotted line represents the average eigenvalues after 100 Monte Carlo simulations shuffling the original data. The zoomed plot shows the significant components (stars) which are over the noise level components (triangles). }
            \label{fig:plotssa}
        \end{figure}

\begin{figure*}
        	\includegraphics[width=\textwidth]{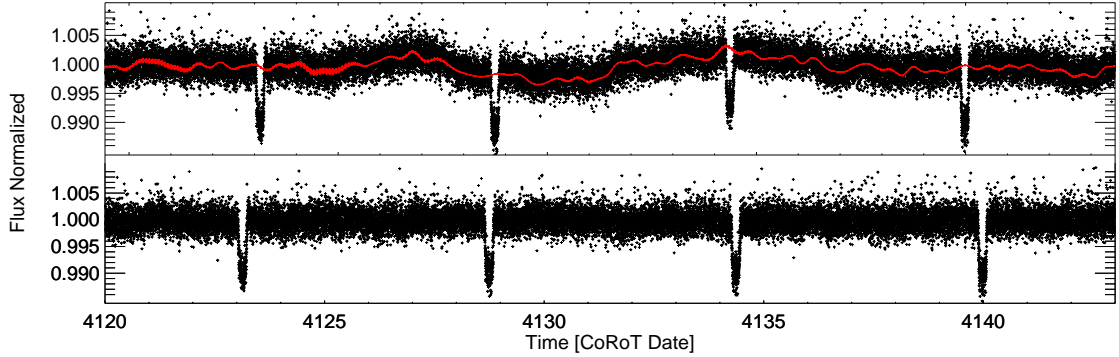}
            \caption{Singular spectrum reconstruction for the CoRoT ID 652345526 light curve. There were used the five eigenvalues that are distinguishable from noise level to reconstruct the variability of the light curve excluding the transits. The top image shows the original light curve and signal reconstruction represented by the black line. At the bottom the transits corrected for the variability.}
            \label{fig:ssamodelplot}
        \end{figure*}

        Using this corrected light curve without transits we did the SSA of the time series and determined the significant components to reconstruct the signal. Doing so we are also able to some degree to remove the variability convolved with the transits and thus reduce the impact of the detrend in the transit parameters. The signal reconstruction from the SSA components was done using the eigenvalues that were above the noise level present in the light curve. To measure which eigenvalues could be distinguished from pure noise we performed a 100 times Monte-Carlo simulation shuffling the data for each light curve as employed before by \cite{emilio2010}. Figure \ref{fig:plotssa} shows the eigenvalues for 250 modes and the average eigenvalues found with the Monte-Carlo simulation. Figure \ref{fig:ssamodelplot} exhibits the signal reconstruction based on the eigenvalues calculated on the previous plot and the light curve detrended with only the transits signals.

\section{Spectroscopic analysis}
\label{sec:spectra}

    \subsection{Measurements}
    \label{sec:measure}

        From the list of host stars from the photometric data suspected or confirmed of having at least one exoplanet, spectroscopic data from the HARPS instrument were retrieved from the \href{http://archive.eso.org}{ESO Science Archive Facility} \citep{Retzlaff14}. HARPS is a fiber-fed cross dispersed spectrograph installed at the coud\'e west focus of the La Silla 3.6m telescope. Its main goal is the search for exoplanets using the radial velocity method with high accuracy (1 m/s) \citep{Mayor03}. Table \ref{tab:spectra} lists the HARPS observations ranging from 2010 to 2013 for the host stars of interest that were obtained.

    \subsection{Stellar and Planetary Parameters}
    \label{sec:stpar}

        Spectra available from the ESO data bank are already reduced and wavelength calibrated, using the HARPS pipeline. \revision{Each spectrum was obtained using the high accuracy mode with one fiber locked on the star and the other on the sky. The radial velocity was measured on each extracted spectrum by means of weighted cross-correlations with a G2 stellar mask.} The uncertainty in RV measurements are mostly due to contributions by photon and readout noise, wavelength calibration error and instrumental drift error \citep{fischer2016}.

\begin{figure*}
        	\includegraphics[angle=0,width=\textwidth]{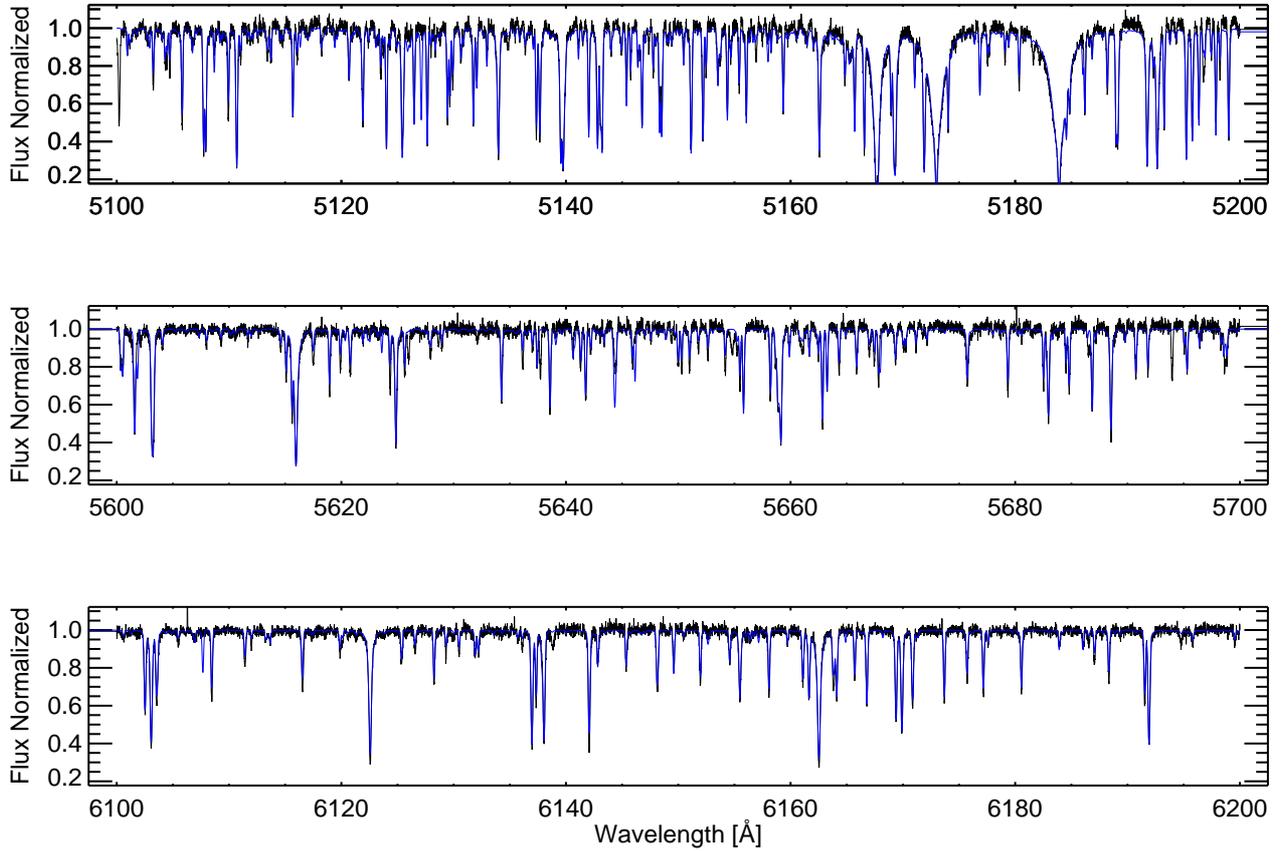}
        	\caption{Spectral synthesis results using the SME algorithm for CoRoT ID 223977153 done with a master spectrum integrated from the HARPS observed spectra. The blue line corresponds to the final synthesis.}
            \label{fig:plotspectrum}
        \end{figure*}

        The spectroscopic analysis has been performed in two steps. First, spectra were analyzed with the Spectroscopy Made Easy (SME) package \citep{valenti1996} that derive accurate stellar parameters (temperature, surface gravity, $\feh$, etc.). SME then produces a spectral synthesis of the star. Figure \ref{fig:plotspectrum} shows the spectral synthesis for CoRoT ID 223977153 (table \ref{tab:stellarparameterstab}).

        Afterwards, $\teff$, $\logg$ and $\feh$ were used as input for the Exofast algorithm \citep{Eastman13}. Exofast is an ensemble of routines that may be used to fit simultaneously exoplanetary photometric transits and radial velocity variations. The code makes a model of the planetary transit taking into account stellar and transit characteristics, giving as output the stellar and the planetary parameters, with the help of accurate mass and radius determination of stars by \citet{Torres10}. We applied the Exofast algorithm to our three stars. The results are given on Tables \ref{tab:stellarparameterstab} and \ref{tab:planetstab}. 

\begin{figure}
        	\includegraphics[angle=0,width=\columnwidth]{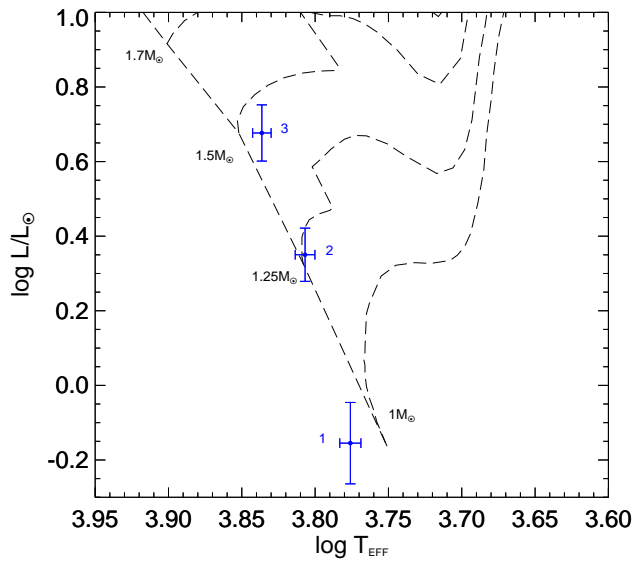}
        	\caption{Evolutionary tracks for selected solar mass stars using the Geneva stellar models \citep{schaller1992}. Crosses represent the parameters of our selected stars derived using the Exofast code. [1] CoRoT ID 223977153; [2] CoRoT ID 104848249 and [3] CoRoT ID 652345526. See table \ref{tab:stellarparameterstab}.}
            \label{fig:hrdiagram}
        \end{figure}

        As can be seen in Fig.~\ref{fig:hrdiagram}, the results for the stellar parameters agree nicely with those expected for stars of similar masses and solar metallicity computed with the Geneva evolutionary track models \citep{schaller1992}.

        %spectroscopic data
        \begin{table}
            \caption{HARPS spectroscopic data retrieved from the \href{http://archive.eso.org}{ESO Science Archive Facility} for the host stars presented.}
            \label{tab:spectra} %
            \begin{tabular}{lcc}
            \hline 
            ~~~CoRoT ID  & Number of spectra  & Interval of observations \tabularnewline
            \hline 
            ~~~223977153 & 21 & 23/01/2010 \textendash{} 05/02/2012 \tabularnewline
            ~~~104848249 & 11 & 17/06/2012 \textendash{} 28/09/2013 \tabularnewline
            ~~~652345526 & 06 & 14/06/2012 \textendash{} 09/07/2012 \tabularnewline
            \hline
            \end{tabular}
        \end{table}
    
        %stellar parameters
        \begin{table*}
            \caption{Stellar Parameters based on calculations provided by the Exofast algorithm.}
            \label{tab:stellarparameterstab} %
            \begin{tabular}{lccccc}
            \hline 
            ~~~Parameter  & Units  &  & ID 223977153  & ID 104848249  & ID 652345526 \tabularnewline
            \hline 
            ~~~$M_{*}$\dotfill{}  & Mass (\msun)\dotfill{}  &  & $1.08_{-0.07}^{+0.08}$  & $1.18_{-0.07}^{+0.08}$  & $1.41\pm0.09$ \tabularnewline
            ~~~$R_{*}$\dotfill{}  & Radius (\rsun)\dotfill{}  &  & $0.79_{-0.09}^{+0.08}$  & $1.22\pm0.09$  & $1.6\pm0.1$ \tabularnewline
            ~~~$L_{*}$\dotfill{}  & Luminosity (\lsun)\dotfill{}  &  & $0.7\pm0.2$  & $2.24\pm0.4$  & $4.75_{-0.8}^{+0.9}$ \tabularnewline
            ~~~$\log(g_{*})$\dotfill{}  & Surface gravity (cgs)\dotfill{}  &  & $4.67_{-0.08}^{+0.11}$  & $4.21_{-0.06}^{+0.07}$  & $4.13_{-0.07}^{+0.05}$ \tabularnewline
            ~~~$\teff$\dotfill{}  & Effective temperature (K)\dotfill{}  &  & $5970\pm100$  & $6410\pm100$  & $6860\pm100$ \tabularnewline
            ~~~$\feh$\dotfill{}  & Metalicity\dotfill{}  &  & $0.0\pm0.2$  & $-0.2\pm0.2$  & $0.0\pm0.2$ \tabularnewline
            ~~~V$sin(i)^\text{a}$\dotfill{} & Rotational Velocity (km/s)\dotfill{} &  & $3.2\pm1.0$ & $23\pm1$ & $21\pm1$ \tabularnewline
            ~~~Spectral Class\dotfill{}&\dotfill{}&  & G0VI  & F7V  & F4V \tabularnewline
            ~~~Magnitude G band\dotfill{} & \dotfill{} &  & $14.108\pm0.003^\text{b}$ & $13.433\pm0.001^\text{c}$ & $12.870\pm0.001^\text{c}$ \tabularnewline 

            \hline 
            \end{tabular}

        \begin{flushleft}
            $^\text{a}$  Rotational velocities were calculated with the SME algorithm. 
        
            $^\text{b}$  Magnitude obtained from SDSS9 catalog \citep{ahn2012}. 

            $^\text{c}$  Magnitude obtained from Gaia catalog \citep{gaia2016a,gaia2016b}.
        \end{flushleft}

        \end{table*}

        %spectral synthesis plot

        %evolutionary tracks plot

\section{Results}
\label{sec:results}

    \begin{figure*}
        \centering
            \includegraphics[width=.33\textwidth]{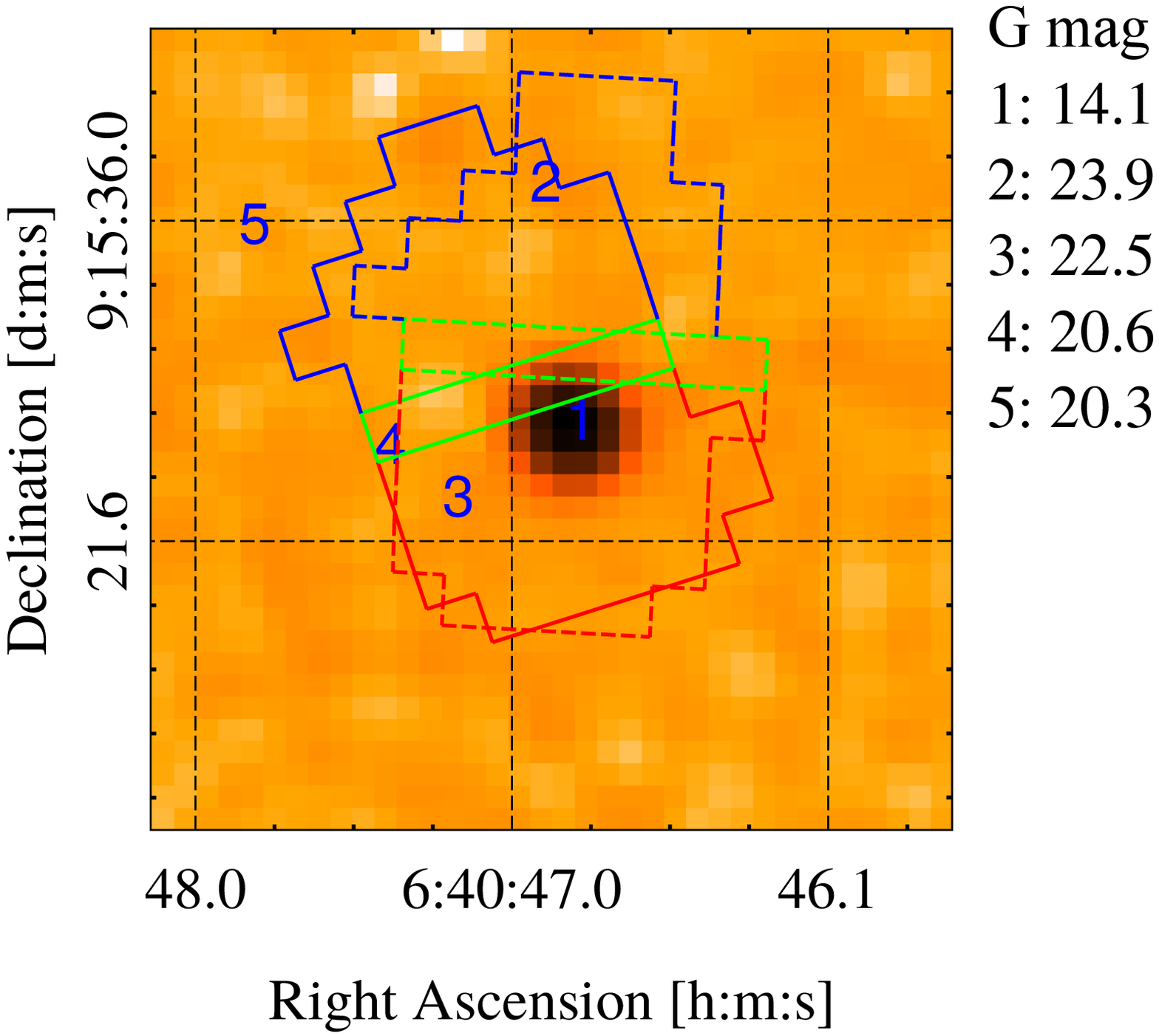}
            \includegraphics[width=.33\textwidth]{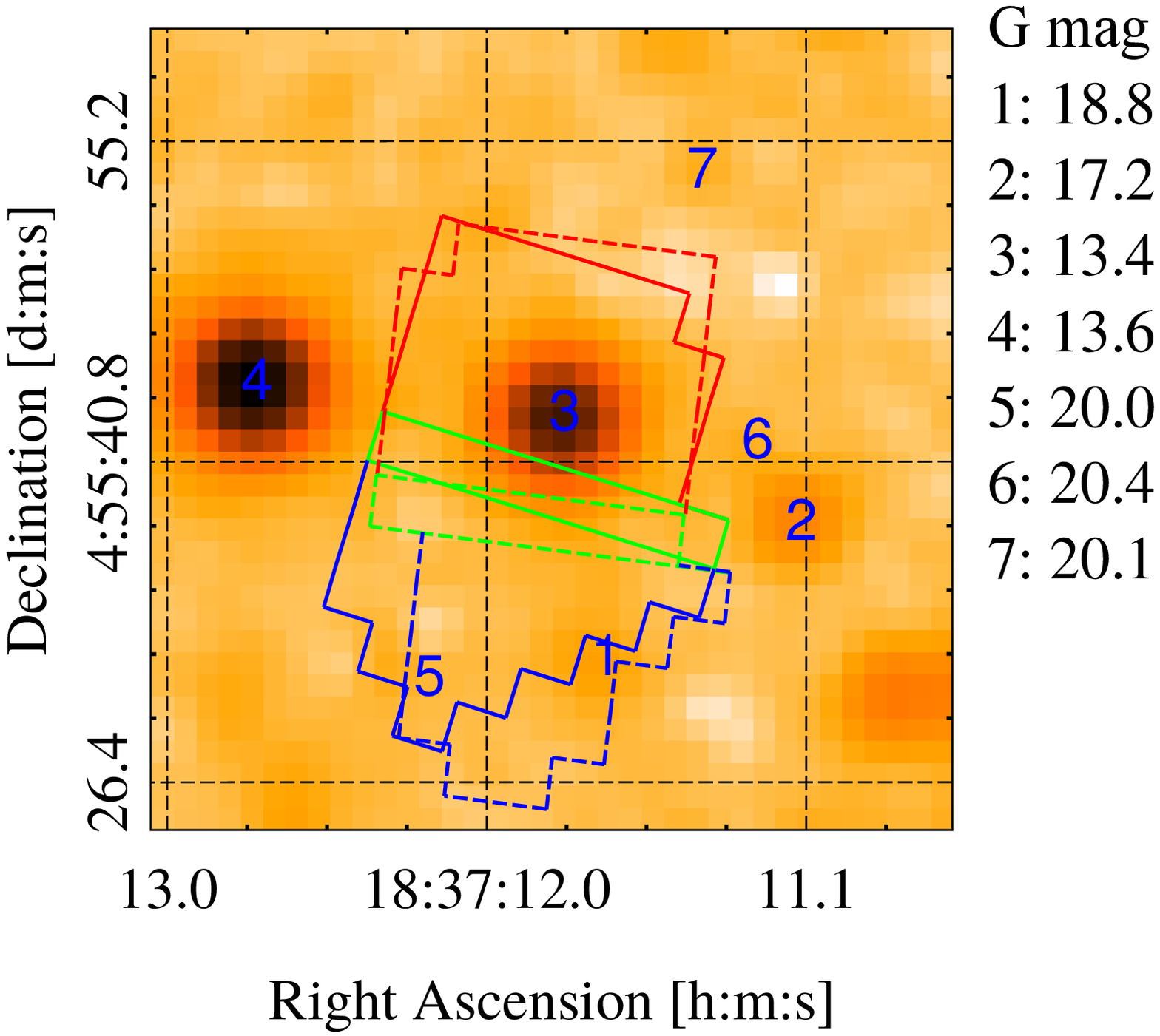}
            \includegraphics[width=.33\textwidth]{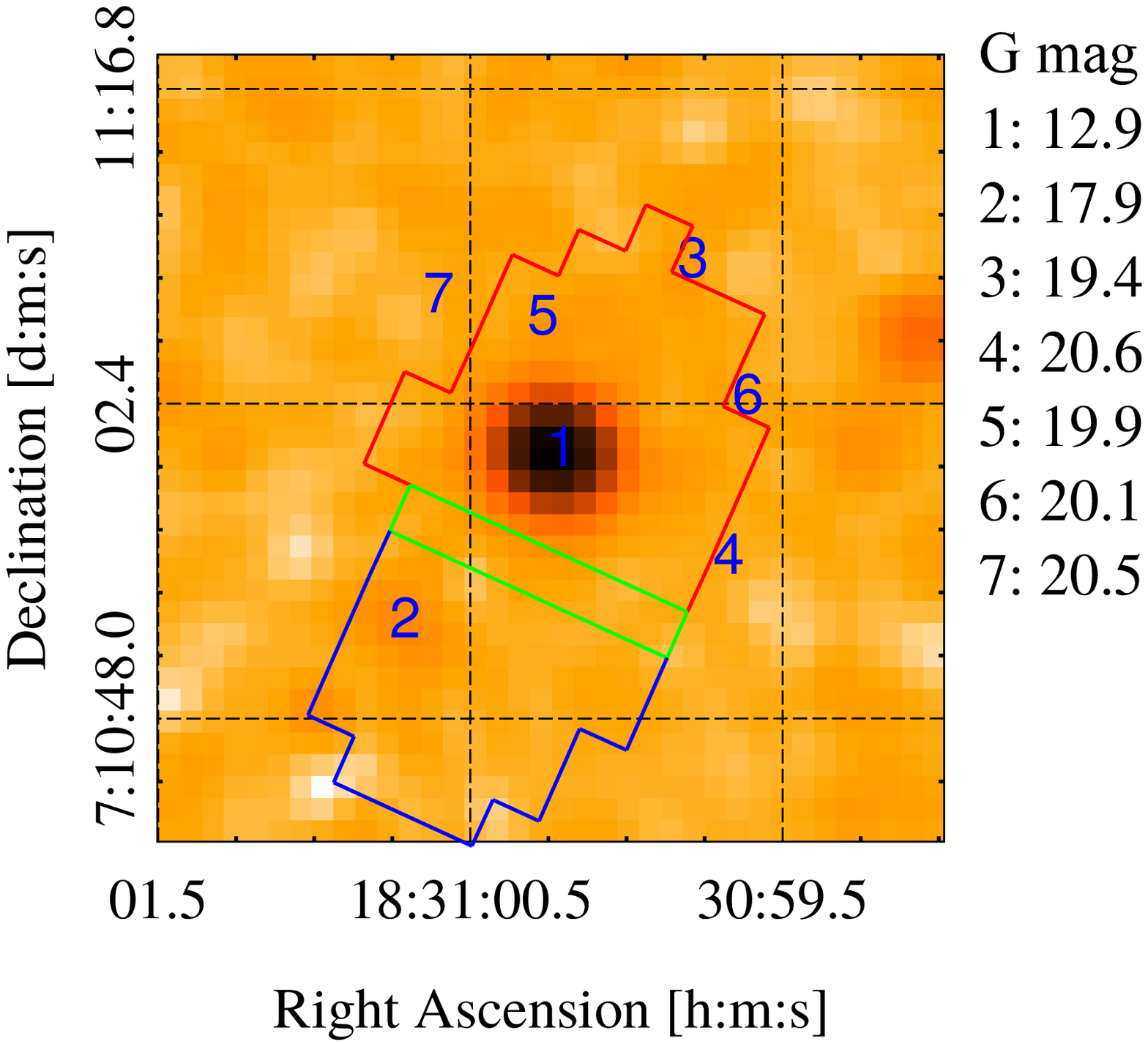}
            \\
            (a) \hspace{57mm} (b) \hspace{57mm} (c)

        \caption{2MASS fields \citep{skrustkie2006} detailing where each target was observed by the CoRoT mission. (a) CoRoT ID 223977153: the target is identified as source 1. A dashed line delineates the mask used during the SRa01 run while the plain line corresponds to the mask at the run SRa05. (b) CoRoT ID 104848249: star marked as source 3. The dashed mask corresponds to the run LRc05 and the plain mask to the run LRc06. (c) Field with CoRoT ID 652345526 and has the star targeted as source 1. It was observed during the run LRc07. Each mask is divided in three regions corresponding to the three channels read out from the CoRoT frames and are indicated with their corresponding colors: red, green and blue. Magnitudes of contamination sources inside and neighbouring the masks are listed at the right side of each image and they were obtained from the SDSS9 (a) and Gaia (b and c) catalogs.}

        \label{fig:background}
    \end{figure*}

    The contamination of eclipsing binaries diluted in the observed targets' flux is one major concern in the analysis of transit candidates and is commonly performed with follow up photometry. Two of our targets (CoRoT IDs 223977153 and 104848249) were observed in two CoRoT runs with different masks. In this work we also made use of the unprecedented data released by the Gaia mission \citep{gaia2016a,gaia2016b} as well as of the Sloan Digital Sky Survey Ninth Release - SDSS9 catalog \citep{ahn2012}. To rule out the sources that might be bright enough to mimic transit features we detailed all known stars present inside each object's observed mask and, taking the spectral response of the CoRoT satellite detectors as reference \citep{levacher2006,auvergne2009}, estimated the total flux contribution of background sources. More details in the following subsections.
    
    The measurements of the radial velocity of our targets where taken in a very irregular time spacing. This introduce false peaks in the power spectrum. We choose then to make the frequency analysis of the radial velocity measurements obtained by HARPS using the CLEANEST Algorithm \citep{foster96} and Period 04 \citep{period04}. CLEANEST is a time analysis technique to detect signals in time series with irregular time spacing. It uses statistics based from Lomb-Scargle  modified periodogram  \citep{scargle82} and the date-compensated discrete Fourier transform \citep{ferraz-mello81}. The power ppectrum follows a chi-square distribution with an expected value of one and two degrees of freedom per frequency fitted. Period 04 is a tool written in Java/C++ to find frequencies in astronomical time series containing gaps. It allows the calculation of the uncertainties of the fitted parameters by means of a Monte Carlo simulation. Both algorithms resulted in consistent results. 
    
    We also performed a F-test. This statistical test if the ratio of the sum of residuals squared decreases significant more than the relative change of the degrees of freedom from a simple model to a more complex one. The planetary system parameters were obtained using the Exofast algorithm \citet{Eastman13} as explained in the previous section.

    \subsection{CoRoT ID 223977153}
        The object CoRoT 223977153-a -- a faint star with coordinates $\alpha = 06$:$40$:$46.843$ and $\delta = +09$:$15$:$26.752$ -- was observed in two runs during the mission. More details on the stellar parameters can be found in Table \ref{tab:stellarparameterstab}. The first measurements were taken in the SRa01 run during March, 2008. The second set was observed during the SRa05 campaign from November, 2011 to January, 2012.
        
        Previous studies had already been carried out for this target. \cite{guenter2013} did high angular resolution measurements and infrared spectroscopy, but did not find any candidate companions very close to the star. \cite{klagyivik2013} performed a variability survey in the CoRoT SRa01 field and classified the star as a Gamma Doradus variable with period of $0.915\pm0.005$ days and amplitude $0.015\pm0.008$ mag. Therefore none of the CoRoT photometric light curves have this period and our SME analysis of the HARPS spectra classify this star as G0VI.

%planet 223977153 photometry
        \begin{figure}
    	    \includegraphics[width=\columnwidth]{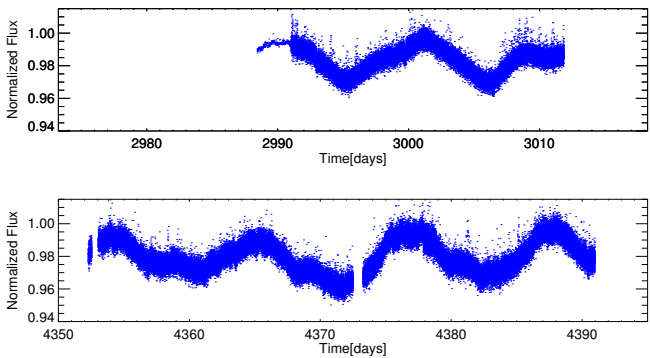}
            \caption{Light curves for CoRoT ID 223977153.Top: Measurements of SRa01 campaign during March, 2008. Bottom: Measurements of SRa05 campaign from November, 2011 to January, 2012.}
            \label{fig:plotfot223977153}
        \end{figure}

        \begin{figure}
        	\includegraphics[width=\columnwidth]{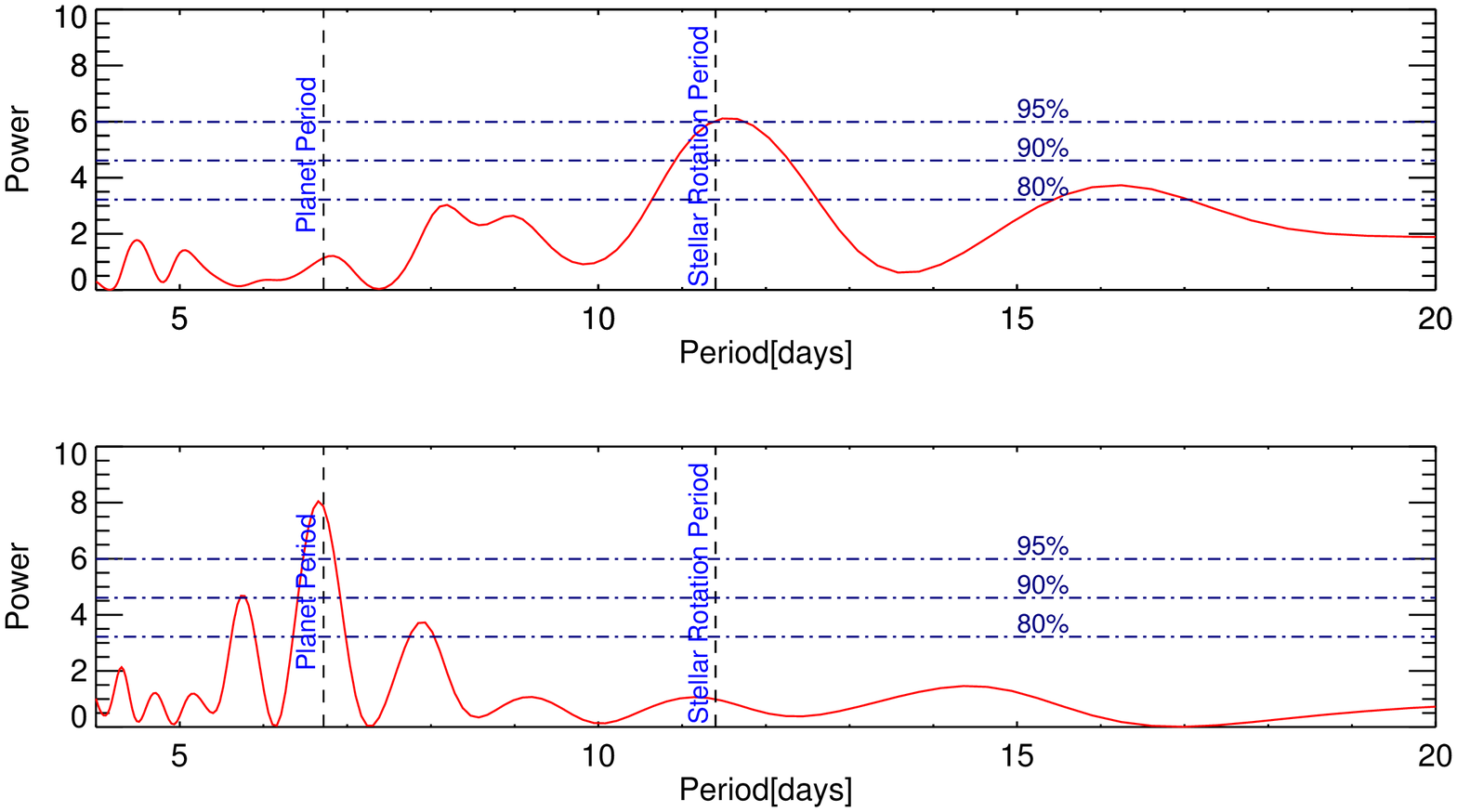}
        	\caption{Top: CLEANEST Spectrum of 21 measurements of radial velocity of CoRoT ID 223977153. The rotation and the Planet Period are show in vertical dashed lines  and the significant levels are shown in dashed horizontal lines. Rotational period appeared as more than $95\%$ of significance in the CLEANEST power spectrum. Bottom: Residual CLEANEST Spectrum after subtracting the rotation signal resulted from our least-square fit from the radial velocity measurements. The same period of $\sim$6.7 days found in the photometric measurements appear as $98\%$ of significance in the CLEANEST power spectrum. }
            \label{fig:clean223977153}
        \end{figure}
        
        To reinforce the analysis of background eclipsing binaries in addition to \cite{guenter2013} we analyzed the transits depth in the three channels separately and compared the depths with all sources from the SDSS9 catalog overlaping with the target's mask. We measured transits' depths of $0.0027\pm0.0004$ (red), $0.0029\pm0.0007$ (green) and $0.0028\pm0.0009$ (blue). Figure \ref{fig:background} (a) shows the contaminating sources present in the mask during the two runs. We converted the magnitudes into relative fluxes and used the transit depth as an estimate of what would be the apparent magnitude contamination limit. In this case, a $0.3$\% flux variation sets a limit of $20.45\pm0.08$ mag. None of the sources present in the mask were bright enough to possibly mimic the observed transit. The total contamination as a function of the target star inside the mask is of $0.06\pm0.02$\%.
        
\begin{figure}
        	\includegraphics[width=\columnwidth]{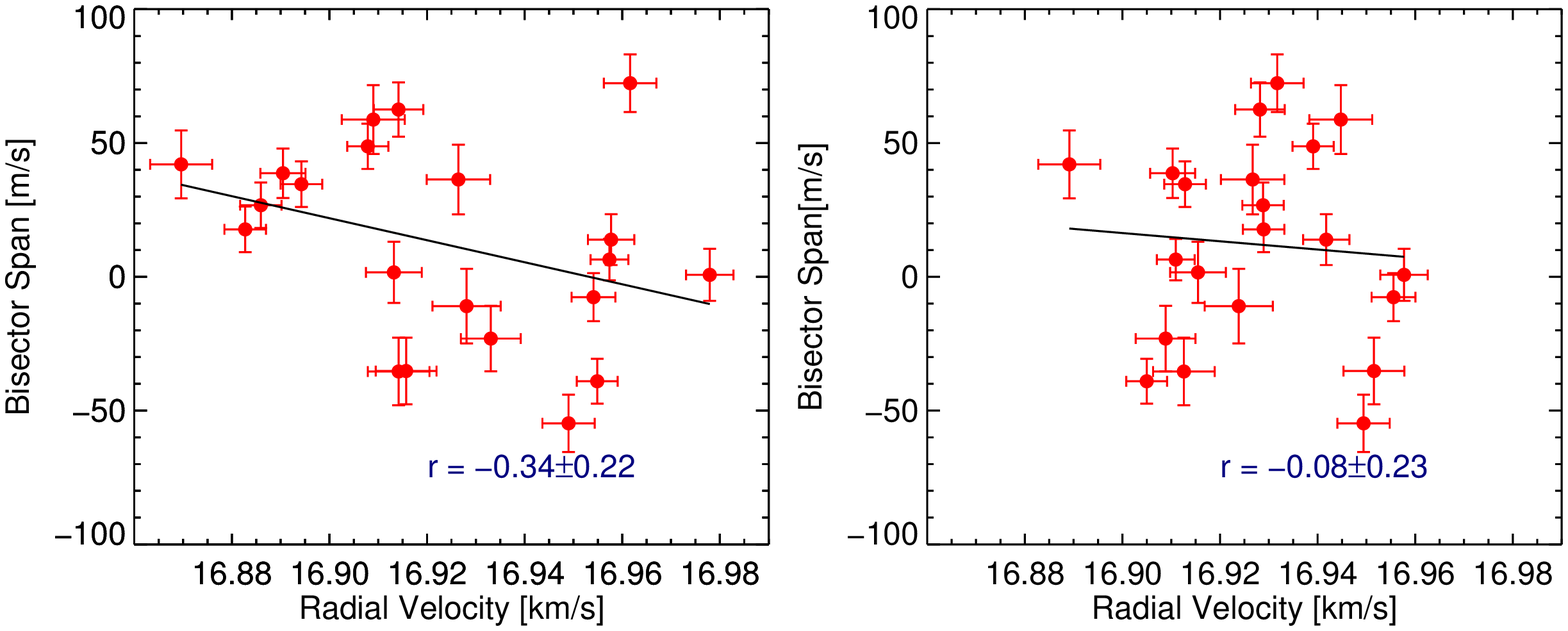}
        	\caption{Left: Anti-correlation between bisector span and the radial velocity measurements of CoRoT ID 223977153 indicating the radial velocity measurements are affected by the stellar activity. Right: The same plot after subtracting from a stellar rotational model.}
            \label{fig:bisec223977153}
        \end{figure}

        This star presents a strong activity visible in the light curves seen in both epochs (see Fig.~\ref{fig:plotfot223977153}). A modulation of $\sim$11.4 days in the photometric data is best explained as due to stellar rotation. The same rotational frequency is found in radial velocity measurements. The $vsin(i)$ found by the SME algorithm resulted in $3.2\pm1.0$ km/s, with the estimated radius for this star, the rotation period (supposing $i=90^{\circ}$) is $14\pm6$ days. Inside the 1$\sigma$ error of the rotation photometric period found.  Fig.~\ref{fig:clean223977153} top panel shows the CLEANEST spectrum of the 21 radial velocity measurements of the CoRoT 223977153 star. The rotational and the planetary periods are shown in vertical dashed lines and the significant levels are shown in dashed horizontal lines. Rotational period appeared as more than $95\%$ of significance in the CLEANEST power spectrum. The precision found in this frequency agrees with the photometric measurements within three significant figures. Another strong indicator is the anti-correlation observed when the bisector span was plotted against the radial velocity measurements (left panel of Fig.~\ref{fig:bisec223977153}). 
        
        We modeled this frequency up to its second harmonic and subtracted it from the radial velocity data as suggested by \cite{boisse2011}, taking into account the phase difference in each data set. The first 8 radial measurements for this star is separated by 692 days from the last 13 measurements. The configuration of the star spots is expected to be different from each set, what will make the amplitude and phase different for the rotation. We notice that when making individual power spectrum for each of the two sets of data. The rotation period found in photometry measurements they also have different phases. The best result came when we least-square fit at once both data sets with the same rotation period but different phases and amplitudes and a unique phase and amplitude for the second period found in the photometric data plus a constant. 
        %A Monte-Carlo simulation were made to estimated the errors of the fitted parameters and to propagate the error of the rotation period. The amplitudes of the fundamental period of the rotation signal are inside the 2$\sigma$ error bar for both data sets but the phases are significant different. The amplitude of the 6.7 day period is $23\pm0.9$ m/s. 
        Fig.~\ref{fig:clean223977153} bottom panel shows the residual CLEANEST spectrum after subtracting the rotation signal resulted from our least-square fit from the radial velocity measurements. The same period of $\sim$6.7 days found in the photometric measurements appear as $98\%$ of significance in the CLEANEST power spectrum. We also subtract the rotation signal from the bisector span plot to show the decrease of the anti-correlation (right panel of Fig.~\ref{fig:bisec223977153}). 
        
        Although the result obtained with the CLEANEST spectrum is statistically significant, we test if it is substantively significant analyzing the spread on the weighted residuals given the hypothesis of the presence of a planet (see \cite{beauge2012}). We ran $10^6$ Monte Carlo realizations to compare the best weighted residual for two models: (a) only rotational modulation due to stellar activity is present in the radial velocity data and (b) rotational modulation plus the gravitational pull of a planet. In this test we used only the last set of the HARPS observations (13 measurements) to maximize the degrees of freedom ruling out the differences in rotational amplitude and phase between the two data sets.

        The analysis took into account the time of the transits and the periods determined from photometry and we used only the fundamental rotational frequency. Before the test was performed, a study on orbital eccentricity was also done and, given the lack of data, it is equally likely to have $e=0$ as values up to $e=0.2$. Thus, we assume a more simple circular orbital model. \revision{The best solution found for the model \emph{a} minimized the residuals (weighted root mean square) to 13.8~m/s. When we compare the results with \emph{b} we see a reduction to 9.4~m/s.} The improvement observed lies close to 2$\sigma$ which would give us a FAP not far from $4.5$\%.
        
\begin{figure}
        	\includegraphics[angle=270,width=\columnwidth]{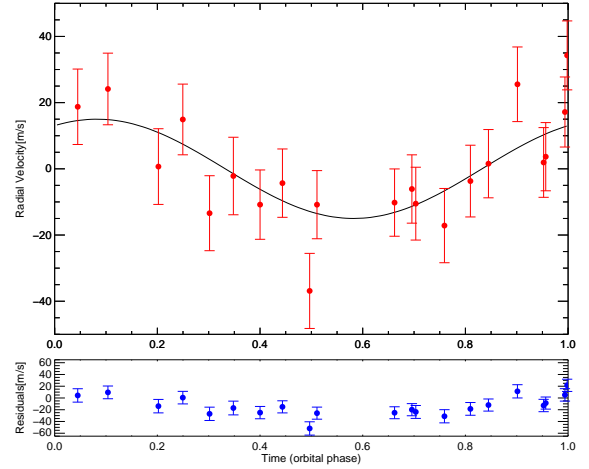}
            \caption{Top: Best solution for the radial velocity measurements of CoRoT ID 223977153 after subtraction of the rotation period and its two first harmonics (see Table \ref{tab:planetstab}). \revision{In contrast to table \ref{tab:223977159rvdata}, the error bars account for the propagation of the uncertainties in the star's rotation model. Bottom: Residuals after subtracting the planet's orbital model.}}
            \label{fig:plotrv223977153}
        \end{figure}

\begin{figure}
    	    \includegraphics[angle=0,width=\columnwidth]{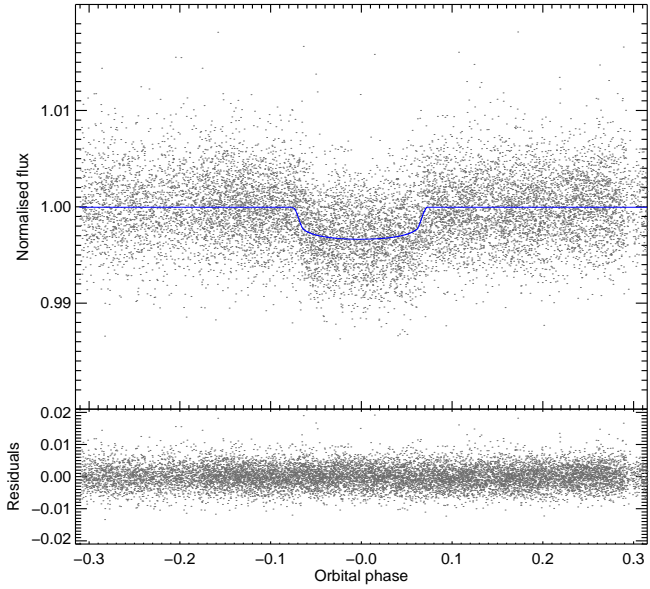}
            \caption{Folded light curve for CoRoT ID 223977153.}
            \label{fig:plot223977153}
        \end{figure}

        \revision{Therefore, to perform the analysis of the planet's signature in the radial velocity data, we propagated in quadrature the uncertainties of the stellar rotation model.} In Figure \ref{fig:plotrv223977153} we show the radial velocity measurements and the best orbital solution for CoRoT ID 223977153. A hot giant planet was found around the star. Table \ref{tab:planetstab} shows the planetary parameters computed with the Exofast algorithm.  The semi-amplitude velocity of the planet found with realistic errors is $14\pm10$~m/s and corresponds to a body with $0.15\pm0.1$\mj~and radius $0.57_{-0.05}^{+0.06}$\rj~with a period of $6.71837\pm0.00001$~days found both in photometric and spectroscopy data. The high precision of the period determination is due to the fact that the two data sets allowed the fitting of the transits within an interval of more than 200 orbits. Figure \ref{fig:plot223977153} shows the folded light curve for CoRoT ID 223977153. The planet orbits its host star which has a mass comparable to our Sun at a semi-major axis smaller than Mercury's orbit.
        
        %4-planet model rv plot
    
        %planet 223977153 folded lc plot

    \subsection{CoRoT ID 104848249}
\begin{figure}
        	\includegraphics[width=\columnwidth]{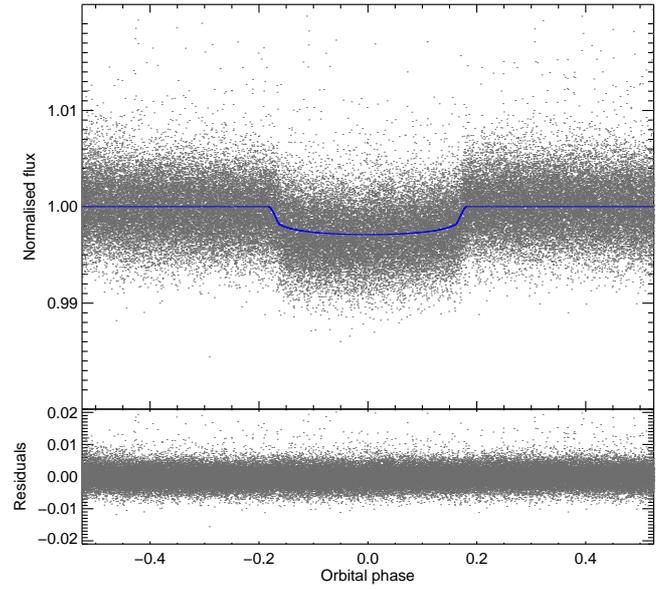}
        	\caption{The folded light curve for CoRoT 104848249.}
            \label{fig:104848249tran}
        \end{figure}

       Figure \ref{fig:104848249tran} shows the folded light curve for the CoRoT ID 104848249 observed during the campaigns LRc05 and LRc06. The star has coordinates $\alpha = 18$:$37$:$11.798$ and $\delta = +04$:$55$:$42.726$ and its apparent magnitude in the R band is $13.7$.  The observational span in photometry is of 168.3 days with a short interval of a few days in between, from April to September, 2010.

       The separate analysis of the transits in the three channels revealed depths of $0.0025\pm0.0001$ (red), $0.0025\pm0.0002$ (green) and $0.0028\pm0.0002$ (blue). Using the Gaia photometry information we estimated the contamination limit for this transit depth in $19.74\pm0.04$ mag in the G band. Figure \ref{fig:background} (b) shows two contaminants inside the mask during the run LRc05 and one contaminant during the run LRc06. Both the cases are, in terms of brightness, close to the edge of the contamination limit as well as located at the edges of the blue channel. A comparison among the transits' depths and their uncertainties strongly suggests that they are actually a feature from the target star leaving little room for the case of a background contamination. The total contamination inside the mask as a function of the target star is estimated in $0.9\pm0.4$\% in the run LRc05 and $0.7\pm0.3$\% in the run LRc06.  
       
       The best fit parameters for the transit model are listed in Table \ref{tab:planetstab}. The candidate has period of $5.70852\pm0.00007$~days and given the stellar parameters we calculate a radius of $0.65\pm0.06$\rj~and the semi-major axis of $0.066\pm0.001$~AU.
       
\begin{figure}
        	\includegraphics[angle=0,width=\columnwidth]{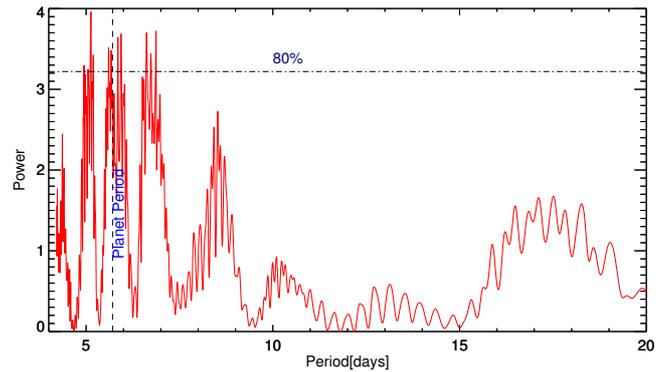}
        	\caption{CLEANEST Spectrum of 11 measurements of radial velocity of CoRoT ID 104848249. The photometric period is shown in vertical dashed lines  and the significant levels are shown in dashed horizontal lines.}
            \label{fig:clean104848249}
        \end{figure}

       Eleven radial velocity measurements were available at ESO Science Archive Facility covering 467.7 days. Fig \ref{fig:clean104848249} shows the CLEANEST power spectrum of the radial velocity measurements for this star. The power spectrum is dominated by the window function. The photometric period of $5.7$ days appear in the CLEANEST spectrum among a group of frequencies where the strongest peak have more than $80\%$ of significance. The maximum semi amplitude span found in the data is around $660$~m/s and could indicate the presence of a companion up to $6$\mj. The absence of anti-correlation between the bisector span and radial velocity suggest no perturbing stellar activity in the radial velocity data. More radial velocity measurements are needed to confirm beyond doubt the nature of the companion.
       
        %planet 104848249 folded lc plot

    \subsection{CoRoT ID 652345526}
        The photometry for this target was taken during the LRc07 run at coordinates $\alpha = 18$:$31$:$00.241$ and $\delta = +07$:$11$:$00.125$. It is a faint star with apparent magnitude in the R band of 13.0 and fifteen transits were observed during this time coverage.     

        The transit analysis in the three colors has shown to be very consistent. We measured a depth of $0.0086\pm0.0006$ in the red channel, $0.009\pm0.001$ in the green channel and $0.008\pm0.001$ in the blue channel. From the Gaia catalog we have a magnitude in the G band of $12.870\pm0.001$, and an estimated contamination limit of $17.97\pm0.04$ mag. Figure \ref{fig:background} (c) shows two contaminants inside the observed mask with one case (object 2) where the measured magnitude overlaps the contamination limit. But the one magnitude level lower uncertainties found in the dispersion of the transit signal in the red channel antagonizes the fact that most of the signal comes from the blue channel. The amount of flux contamination inside the mask as a function of the target star is estimated in $1.0\pm0.6$\%.
        
\begin{figure}
        	\includegraphics[width=\columnwidth]{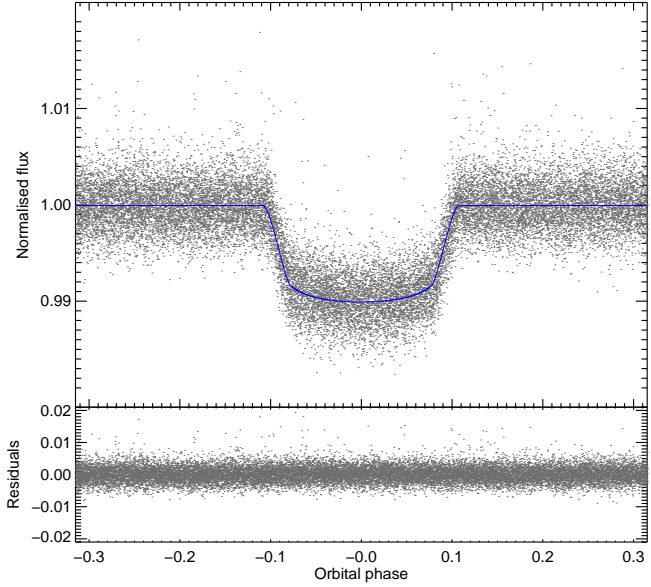}
        	\caption{The folded light curve for CoRoT 652345526.}
            \label{fig:652345526tran}
        \end{figure}

        The period found for this candidate is of $5.61618\pm0.00004$~days. It has $1.4\pm0.1$\rj~and orbits the star at a distance of $0.069\pm0.002$~AU. Figure~\ref{fig:652345526tran} shows the fold phased light curve and the residual from the adjusted model. More information about the transit and companion parameters are given in Table~\ref{tab:planetstab}. Only six radial velocity measurements were available at the ESO Science Archive Facility covering $21.9$~days for this star. We could make no further conclusions on the nature of the candidate. Moreover, we did not see strong indications of stellar activity given the stellar rotational velocity.

        %planet 652345526 plot

%planetary parameters for 223977153, 104848249 and 652345526
\begin{table*}
\caption{Planetary parameters for the systems CoRoT ID 223977153, CoRoT ID
104848249 and CoRoT ID 652345526. The outputs were obtained with the
Exofast algorithm.}
\label{tab:planetstab} %
\begin{tabular}{lcccc}
\hline 
~~~Parameter  & Units  & 223977153-b  & 104848249 cand.  & 652345526 cand. \tabularnewline
\hline 
Planetary Parameters:  &  &  &  & \tabularnewline
~~~$P$\dotfill{} & Period (days)\dotfill{} & $6.71837\pm0.00001$  & $5.70852\pm0.00007$  & $5.61618\pm0.00004$ \tabularnewline
%~~~$e$\dotfill{} & Eccentricity\dotfill{} & $0.17\pm0.08$  &  & \tabularnewline
%~~~$\omega_{*}$\dotfill{} & Argument of periastron (days)\dotfill{} & $-12_{-26}^{+29}$  &  & \tabularnewline
~~~$a$\dotfill{} & Semi-major axis (AU)\dotfill{} & $0.071\pm0.001$  & $0.066\pm0.001$  & $0.069\pm0.002$ \tabularnewline
~~~$M_{P}$\dotfill{} & Mass (\mj)\dotfill{} & $0.15\pm0.10$  &              &           \tabularnewline
~~~$R_{P}$\dotfill{} & Radius (\rj)\dotfill{} & $0.57_{-0.05}^{+0.06}$  & $0.65\pm0.06$  & $1.4\pm0.1$ \tabularnewline
~~~$\rho_{P}$\dotfill{} & Density (cgs)\dotfill{} & $1.1_{-0.8}^{+1.3}$  &   &  \tabularnewline
~~~$\log(g_{P})$\dotfill{} & Surface gravity\dotfill{} & $3.1_{-0.3}^{+0.6}$  & & \tabularnewline
~~~$T_{eq}$\dotfill{} & Equilibrium Temperature (K)\dotfill{} & $938_{-52}^{+65}$  & $1330\pm49$  & $1560\pm65$ \tabularnewline
~~~$\fave$\dotfill{} & Incident flux (\fluxcgs)\dotfill{} & $0.18_{-0.04}^{+0.05}$  & $0.7\pm0.1$  & $1.4\pm0.2$ \tabularnewline
RV Parameters:  &  &  &  & \tabularnewline
~~~$K$\dotfill{} & RV semi-amplitude (m/s)\dotfill{} & $14\pm10$  &  &  \tabularnewline
%~~~$T_{P}$\dotfill{} & Time of periastron (\bjdtdb)\dotfill{} & $2455766.4\pm0.5$  &  & \tabularnewline
%~~~$M_{P}/M_{*}$\dotfill{} & Mass ratio\dotfill{} & $0.00018\pm0.00002$  &  & \tabularnewline
Primary Transit Parameters:  &  &  &  & \tabularnewline
~~~$T_{C}$\dotfill{} & Time of transit (\bjdtdb)\dotfill{} & $2455901.811_{-0.001}^{+0.002}$  & $2455299.387\pm0.001$  & $2455662.5203\pm0.0004$ \tabularnewline
~~~$R_{P}/R_{*}$\dotfill{} & Radius of planet in stellar radii\dotfill{} & $0.054\pm0.002$  & $0.0552\pm0.0009$  & $0.095\pm0.002$ \tabularnewline
~~~$u_{1}$\dotfill{} & linear limb-darkening coeff\dotfill{} & $0.35\pm0.05$  & $0.29\pm0.05$  & $0.22\pm0.05$ \tabularnewline
~~~$u_{2}$\dotfill{} & quadratic limb-darkening coeff\dotfill{} & $0.28\pm0.05$  & $0.29\pm0.05$  & $0.30\pm0.05$ \tabularnewline
~~~$i$\dotfill{} & Inclination (degrees)\dotfill{} & $89.0\pm0.5$  & $87.9\pm0.2$  & $87.4_{-0.7}^{1.0}$ \tabularnewline
~~~$b$\dotfill{} & Impact Parameter\dotfill{} & $0.4_{-0.2}^{+0.1}$  & $0.43\pm0.02$  & $0.44_{-0.14}^{+0.08}$ \tabularnewline
~~~$\delta$\dotfill{} & Transit depth\dotfill{} & $0.0029\pm0.0002$  & $0.0030\pm0.0001$  & $0.0091\pm0.0003$ \tabularnewline
~~~$\tau$\dotfill{} & Ingress/egress duration (days)\dotfill{} & $0.006\pm0.001$  & $0.010\pm0.001$  & $0.020\pm0.003$ \tabularnewline
~~~$T_{14}$\dotfill{} & Total duration (days)\dotfill{} & $0.104_{-0.007}^{+0.009}$  & $0.151\pm0.009$  & $0.187\pm0.006$ \tabularnewline
Secondary Eclipse Parameters:  &  &  &  & \tabularnewline
~~~$T_{S}$\dotfill{} & Time of eclipse (\bjdtdb)\dotfill{} & $2455905.170_{-0.001}^{+0.002}$  & $2455302.240\pm0.001$  & $2455665.3284\pm0.0004$ \tabularnewline
\hline 
\end{tabular}
\end{table*}

\section{Conclusions}
\label{sec:conclusions}
    We presented in this work a modification of the CoRoT Detrend Algorithm. The implementation using robust statistics decreased the systematic light curve variations and improved the detection of exoplanets when comparing with the original algorithm. All CHR CoRoT light curves (a total of 65,655) were analyzed in this work with our algorithm. All CoRoT known exoplanets in those light curves were found and we presented the analysis of three CoRoT photometric candidates. Spectroscopic measurements of the host star candidates were retrieved from the ESO Science Archive Facility public data base. We then derived the stellar parameters using the Spectroscopy Made Easy package and the Exofast algorithm.
    CoRoT ID 104848249 candidate has a period of $5.70852\pm0.00007$ and orbits a dwarf star with $1.22\pm0.09$\rsun. CoRoT ID 652345526 candidate has a period of $5.61618\pm0.00004$ days and orbits a F4V star with $1.6\pm0.1$\rsun.  Both CoRoT ID 104848249 and CoRoT ID 652345526 need more radial velocity measurements to confirm the nature of these transits. A hot giant planet was found around the active star CoRoT ID 223977153. With period of $6.71837\pm0.00001$ it is an exoplanet with mass of $0.15\pm0.10$\mj~and $0.57_{-0.05}^{+0.06}$\rj. This star was observed in two CoRoT runs: SRa01 and SRa05.

\section*{Acknowledgements}
    Based on observations collected at the European Organisation for Astronomical Research in the Southern Hemisphere under ESO programmes 184.C-0639(A) and 188.C-0779(A).
    
    This research has made use of the ExoDat Database, operated at LAM-OAMP, Marseille, France, on behalf of the CoRoT/Exoplanet program.
        
    This work was supported by CAPES. We wish to thank the CoRoT team for the acquisition and reduction of the CoRoT data. We thank D. Mislis for the productive discussion during the 2nd CoRoT Symposium in Marseille.

%%%%%%%%%%%%%%%%%%%%%%%%%%%%%%%%%%%%%%%%%%%%%%%%%%



\bibliographystyle{mnras}
\bibliography{otherrefs}
%\bibliography{boufleur2017bibliography} % if your bibtex file is called example.bib

%%%%%%%%%%%%%%%%%%%%%%%%%%%%%%%%%%%%%%%%%%%%%%%%%%

%%%%%%%%%%%%%%%%% APPENDICES %%%%%%%%%%%%%%%%%%%%%

\appendix

\section{Radial velocity data}

%CoRoT ID 223977153 radial velocity data
\begin{table}
    \caption{CoRoT ID 223977153 radial velocities, their errors, and bisector spans.}
    \label{tab:223977159rvdata} 
    \begin{tabular}{|c|c|c|c|}
    \hline
    BJD & RV & Error & Bis \tabularnewline 
        & [km s$^{-1}$] & [km s$^{-1}$] & [km s$^{-1}$] \tabularnewline
    \hline
    2455219.59065 &    16.913 &     0.006 &     0.002 \tabularnewline
    2455223.61076 &    16.914 &     0.006 &    -0.035 \tabularnewline
    2455226.68420 &    16.933 &     0.006 &    -0.023 \tabularnewline
    2455227.63942 &    16.916 &     0.006 &    -0.035 \tabularnewline
    2455228.60403 &    16.909 &     0.006 &     0.059 \tabularnewline
    2455229.65977 &    16.926 &     0.007 &     0.036 \tabularnewline
    2455230.64074 &    16.928 &     0.007 &    -0.011 \tabularnewline
    2455231.63879 &    16.870 &     0.006 &     0.042 \tabularnewline
    2455923.72685 &    16.955 &     0.004 &    -0.039 \tabularnewline
    2455924.74179 &    16.957 &     0.004 &     0.006 \tabularnewline
    2455925.73421 &    16.962 &     0.005 &     0.072 \tabularnewline
    2455926.69239 &    16.958 &     0.005 &     0.014 \tabularnewline
    2455927.70976 &    16.949 &     0.005 &    -0.055 \tabularnewline
    2455928.68951 &    16.914 &     0.005 &     0.063 \tabularnewline
    2455929.70052 &    16.890 &     0.005 &     0.039 \tabularnewline
    2455931.68719 &    16.883 &     0.004 &     0.018 \tabularnewline
    2455932.68897 &    16.886 &     0.004 &     0.027 \tabularnewline
    2455933.71877 &    16.954 &     0.004 &    -0.008 \tabularnewline
    2455953.59364 &    16.908 &     0.004 &     0.049 \tabularnewline
    2455960.56329 &    16.978 &     0.005 &     0.001 \tabularnewline
    2455963.58491 &    16.894 &     0.004 &     0.035 \tabularnewline
    \hline
    \end{tabular}
\end{table}

%CoRoT ID 104848249 radial velocity data
\begin{table}
	\caption{CoRoT ID 104848249 radial velocities, their errors, and bisector spans.}
	\label{tab:104848249rvdata} 
	\begin{tabular}{|c|c|c|c|}
	\hline
	BJD & RV & Error & Bis \tabularnewline 
	    & [km s$^{-1}$] & [km s$^{-1}$] & [km s$^{-1}$] \tabularnewline
	\hline
	2456096.79074 &     3.062 &     0.004 &    -0.514 \tabularnewline
	2456099.76491 &     3.109 &     0.005 &    -4.237 \tabularnewline
	2456117.56291 &     3.036 &     0.004 &    -0.389 \tabularnewline
	2456119.63506 &     3.551 &     0.005 &    -0.160 \tabularnewline
	2456151.65012 &     2.991 &     0.005 &    -0.443 \tabularnewline
	2456154.59511 &     3.652 &     0.005 &    -0.173 \tabularnewline
	2456158.47338 &     3.131 &     0.005 &    -0.010 \tabularnewline
	2456160.56214 &     3.659 &     0.006 &    -1.077 \tabularnewline
	2456509.64282 &     3.271 &     0.005 &     1.163 \tabularnewline
	2456539.53156 &     3.357 &     0.005 &    26.245 \tabularnewline
	2456564.49967 &     3.387 &     0.004 &    -1.155 \tabularnewline
	\hline
	\end{tabular}
\end{table}


%%%%%%%%%%%%%%%%%%%%%%%%%%%%%%%%%%%%%%%%%%%%%%%%%%

% Don't change these lines
\bsp	% typesetting comment
\label{lastpage}

\end{document}